\documentstyle[amssymb,prl,aps,multicol,epsf]{revtex}
\renewcommand{\narrowtext}{\begin{multicols}{2}
\global\columnwidth20.5pc} 
\renewcommand{\widetext}{\end{multicols}
\global\columnwidth42.5pc} \multicolsep = 8pt plus 4pt minus 3pt

\begin{document}
\draft
\title{Intersubband plasmons in quasi-one-dimensional electron systems on a liquid
helium surface}
\author{Marcos R. S. Tavares}
\address{Faculdade de Tecnologia da Baixada Santista, CEETPS, 11045-908 Santos, SP,
Brazil}
\author{G.-Q. Hai}
\address{Instituto de F\'{i}sica de S\~{a}o Carlos, Unversidade de S\~{a}o Paulo,
13560-970 S\~{a}o Carlos, SP, Brazil}
\author{F. M. Peeters}
\address{Department of Physics, University of Antwerp (UIA), B-2610 Antwerpen, Belgium}
\author{Nelson Studart}
\address{Departamento de F\'{i}sica, Universidade Federal de S\~{a}o
Carlos,13565-905, S\~{a}o Carlos, SP, Brazil}
\maketitle

\begin{abstract}
The collective excitation spectra are studied for a multisubband
quasi-one-dimensional electron gas on the surface of liquid helium.
Different intersubband plasmon modes are identified by calculating the
spectral weight function of the electron gas within a 12 subband model.
Strong intersubband coupling and depolarization shifts are found. When the
plasmon energy is close to the energy differences between two subbands,
Landau damping in this finite temperature system leads to plasmon gaps at
small wavevectors.
\end{abstract}

\pacs{}

\narrowtext
The electron system on the surface of liquid helium discovered about 30
years ago has provided an ideal platform to study many-body effects in low
dimensions.\cite{Cole74,GA76} Recently, quasi-one-dimensional (Q1D) electron
gas embedded in such structures has been realized in laboratory by bending
the liquid helium surface.\cite{Kovdrya92} In some sense, this Q1D system is
similar to those in semiconductor quantum wires, nanotubes and metallic
chains where many-body effects have been extensively studied. \cite{Voit}
However, this new system on the surface of liquid helium provides us more
degrees of freedom to explore many-body effects in a 1D framework. It is
free from impurities being much clearer than other systems. Furthermore, it
is of a wider variable range of electron densities. Several theoretical and
experimental studies have been carried out on the electron transport and
electron-electron interactions in this system.\cite{kovrev,Lea01} The
electron system on the liquid helium surface is achieved at finite
temperatures and the electron density is usually much smaller than that in
semiconductor structures. In most cases, it is considered as a
non-degenerate electron gas obeying the Maxwell-Boltzmann statistics.\cite
{SHS95} On the other hand, the 1D confinement is weak and the gap between
the 1D subbands is of the same order as the thermal energy. As consequence,
many subbands are usually occupied forming a real multisubband Q1D electron
system. Very recently, dispersion relation of the collective excitations
(plasmons) was studied theoretically within the random-phase approximation
(RPA) for the Q1D electron gases on the surface of liquid helium.\cite
{SHS95,ss97} The RPA is believed to be a more reliable approximation to
study collective excitations of such a non-degenerate Q1D electron gas in
the classical regime. In these studies,\cite{ss97} a two-subband model was
used and particular attention was devoted to the dispersion relations of one
intrasubband and one intersubband plasmon modes found within this model.
Notice that, both the RPA and the quasi-crystalline approximation, being
valid in opposite limits of electron density at fixed temperature, result in
very similar plasmon dispersions indicating that the RPA could be correct
over a wide range of electron densities.

This Rapid Communication is focused on the collective excitation spectra of
the multisubband Q1D electron gas on the surface of liquid helium. Special
attention is paid to the intersubband coupling and effects of the
single-particle excitations (SPE), {\it i.e.} the Landau damping, on the
plasmon modes at finite temperature. We study these elementary excitations
by calculating the full spectral weight function of this classical electron
system. The Landau damping induced phenomena are expected to be of a new
kind provided the electrons are in a very different regime from those
embedded in semiconductor quantum wires. In a Q1D electron system, the
intersubband interactions are much stronger than those in higher dimensions.
Furthermore, temperature effects are important because the energy gap
between the 1D electron subbands is comparable to the thermal energy $k_{B}T$
. Many subbands can be occupied even for small densities and temperatures.
As a consequence, the intersubband coupling should be treated properly. In
the calculations, we take into account as many subbands as possible to
guarantee the efficiency of the method and to make sure the obtained plasmon
spectra being independent of the number of subbands included.

We consider the same structure for the Q1D electron system as in Ref.\cite
{SHS95} where the electron mobility was studied. The 1D confinement on the
liquid helium is determined by both the surface curvature radius in the $y$%
-direction $R=5\times 10^{-4}$ cm and the so-called holding field applied in
the $z$-direction $E_{\perp }=3\times 10^{3}$ V/cm. This confinement is then
approximated by a parabolic potential in the $y$-direction with a
confinement frequency $\omega _{0}=\sqrt{eE_{\perp }/mR}$, where $m$ and $e$
are the mass and the charge of the electron, respectively. The energy
eigenvalues of an electron in the system is given by $E_{n}(k_{x})=$ $%
k_{x}^{2}/2m$ + $(n-\frac{1}{2})\omega _{0}$, where $k_{x}$ is the electron
wavevector in the $x$-direction and $(n-\frac{1}{2})\omega _{0}$ is the
electron energy levels due to confinement in the $y$-direction, with $n=1,2,$
..., being the subband index. Here, we consider the electron gas being of
zero-thickness in the $z$-direction because the energy gap between the two
lowest levels due to confinement in this direction is greater than 30 K,\cite
{7} whereas the confinement energy $\omega _{0}$ in the $y$-direction is
less than 1 K. Another important parameter is the localization length $%
y_{0}=k_{0}^{-1}=\sqrt{1/2m\omega _{0}}$ of the electrons in the $y$%
-direction (we consider $\hslash =1$ throughout this paper).

Our calculations show that there is only one observable intrasubband plasmon
mode but many intersubband modes in the present Q1D electron system. All the
observable intersubband plasmon modes are related to the first and second
subbands denoted by $(1,n)$ and $(2,n)$, respectively, where $n=1,2,3,...$.
Such intersubband modes are important since most of the electrons occupy
states in the two lowest subbands and intersubband interactions are strong
in a one-dimensional geometry. Furthermore, we find that intersubband
coupling between the higher subbands and the two lowest subbands affects
significantly the low-energy intersubband plasmon modes. A correct
theoretical consideration on such a coupling turns out to be essential in
obtaining even the lowest intersubband plasmon modes. In order to consider
this effect properly, we use in this work a 12 subband model so that the
plasmon spectra shown below do not change anymore when more subbands are
included. Moreover, we show that pronounced Landau damping occurs when the
plasmon energy approaches the energy difference $n\omega _{0}$ between two
subbands leading to gaps in the plasmon spectra.

A traditional way to obtain the plasmon dispersion relations in Q1D electron
gases at zero temperature \cite{JPC} is to find the roots of the equation $%
\det \left| 
%TCIMACRO{\func{Re}}%
%BeginExpansion
\mathop{\rm Re}%
%EndExpansion
\{\varepsilon _{\alpha \beta }(q,\omega )\}\right| =0,$ where $\varepsilon
_{\alpha \beta }$ is the multisubband dielectric matrix and the indices $%
\alpha \equiv (i,i^{\prime })$ and$\ \beta \equiv (j,j^{\prime })$ with the
subband indices $i,i^{\prime },j$ and $j^{\prime }$. However, this method
cannot provide us complete information of the plasmon excitations in the
present system. There are two main reasons for that: (i) it is difficult to
figure out where the plasmon excitations are Landau damped by SPEs because $%
%TCIMACRO{\func{Im}}%
%BeginExpansion
\mathop{\rm Im}%
%EndExpansion
\{\varepsilon _{\alpha \beta }(q,\omega )\}\neq 0$ in the whole $\omega $-$q$
plane at finite temperatures; and (ii) the thermal fluctuations in the
system might easily populate several subbands even for small densities.
Therefore, intersubband interactions can be strong so that a reasonable
amount of subbands should be included in the calculation. The equation $\det
\left| 
%TCIMACRO{\func{Re}}%
%BeginExpansion
\mathop{\rm Re}%
%EndExpansion
\{\varepsilon _{\alpha \beta }\}\right| =0$ yields many roots but does not
provide information of the relative importance of each one. Thus, one cannot
distinguish which roots correspond to the plasmon modes.

A more reliable way to study the collective excitations in such a
multisubband system at finite temperatures is to calculate the so-called
spectral weight \cite{mahan} 
\begin{equation}
S(q,\omega )=-\sum_{\alpha \beta }{\rm Im}\left[ \varepsilon _{\alpha \beta
}^{-1}(q,\omega )\Pi _{\beta }^{\delta }(q,\omega )\right] .  \label{s(qw)}
\end{equation}
The peaks of this function give information of the plasmon excitations and,
from their position, we can obtain the dispersion relations of the plasmon
modes. By showing all excitation modes through $S(q,\omega )$, one provides
a very efficient guide of what should be observable in the experiments.
These observations are certainly dependent on external probes (e.g., light
polarization). Eq. \ref{s(qw)} also involves the 1D non-interacting
irreducible polarizability function $\Pi _{\beta }^{\delta }(q,\omega )$. We
remember that the Maxwell-Boltzmann distribution function is used in
calculating $\Pi _{\beta }^{\delta }(q,\omega )$, with $\delta $ being a
phenomenological constant which is responsible for broadening of the energy
levels $E_{n}(k_{x})$ mainly due to ripplon and evaporated helium atom
scattering on the surface. The quantity $S(q,\omega )$ is directly related
to the optical (such as inelastic light scattering spectra) and transport
(such as conductivity) properties. The multisubband dielectric matrix
function 
\[
\varepsilon _{\alpha \beta }(q,\omega )=\delta _{\alpha \beta }-V_{\alpha
\beta }(q)\cdot \Pi _{\beta }^{\delta }(q,\omega ) 
\]
is written within the RPA, with $V_{\alpha \beta }(q)$ being the Coulomb
electron-electron bare interaction.

The symmetric confinement potential in the $y$-direction leads to the
electron-electron Coulomb interaction $V_{\alpha \beta }(q)=0$ when $%
i+i^{\prime }+j+j^{\prime }$ is an odd number. Consequently, the dielectric
matrix elements (both the real and the imaginary parts) $\varepsilon
_{\alpha \beta }(q,\omega )=0$ for $i+i^{\prime }+j+j^{\prime }={\rm odd}$.
The dielectric matrix can then be decoupled into two submatrices $%
\varepsilon _{\alpha \beta }^{{\rm even}}(q,\omega )$ and $\varepsilon
_{\alpha \beta }^{{\rm odd}}(q,\omega )$ with both $i+i^{\prime }$ and $%
j+j^{\prime }$ being even and odd numbers, respectively. The even (odd)
dielectric submatrix involves only intersubband electron-electron
interaction for one electron from subband $i$ to subband $i^{\prime }$ with $%
i+i^{\prime }={\rm even}$ ($i+i^{\prime }={\rm odd}$) and the other from
subband $j$ to $j%
%TCIMACRO{\UNICODE[m]{0xb4}}%
%BeginExpansion
{\acute{}}%
%EndExpansion
$ with $j+j^{\prime }={\rm even}$ ($j+j^{\prime }={\rm odd).}$ As a
consequence, the spectral weight function can be treated separately in two
parts $S(q,\omega )=S^{{\rm even}}(q,\omega )+S^{{\rm odd}}(q,\omega )$.

In order to understand the damping induced phenomena in the present system,
we calculate first the spectral weight of the single-particle excitations $%
S_{sp}(q,\omega )=-\sum_{jj^{\prime }}{\rm Im}\left[ \Pi _{jj^{\prime
}}^{\delta }(q,\omega )\right] $. It is straightforward to analytically
obtain the imaginary part of the polarizability $\Pi _{\beta }^{\delta
}(q,\omega )$ at finite temperatures.\cite{ss97} Considering the level
broadening effects, it is given by
\[
%TCIMACRO{\func{Im}}%
%BeginExpansion
\mathop{\rm Im}%
%EndExpansion
\left[ \Pi _{ij}^{\delta }\right] =N_{e}\left\{ \exp [-j/T]\text{ }I^{\delta
}\left( \xi _{ij}^{(+)}\right) -\right. 
\]
\begin{equation}
\text{ \ \ }\left. \exp [-i/T]\text{ }I^{\delta }\left( \xi
_{ij}^{(-)}\right) \right\} /q\sqrt{\pi T}\left[ 1+\coth \left( 1/2T\right) %
\right]   \label{pol}
\end{equation}
where $N_{e}$ is the total electron density, $T$ the temperature, $q$ the
wavevector in the $x$-direction, and 
\begin{equation}
I^{\delta }(\xi )=\Delta \cdot \int_{-\infty }^{\infty }dk_{x}\frac{\exp %
\left[ -(k_{x}+\xi )^{2}\right] }{k_{x}^{2}+\Delta ^{2}},  \label{inte}
\end{equation}
with $\xi _{ij}^{(\pm )}=\left[ \omega +(i-j)\pm q^{2}\right] /2q\sqrt{T}\,$%
and $\Delta =\delta /2q\sqrt{T}.$ The wavevectors $k_{x}$ and $q$ are
written in unit $k_{0}$, while the frequency $\omega $, the broadening
constant $\delta $ and the temperature $T$ are in unit $\omega _{0}$. Notice
that, in the limit $\delta \rightarrow 0,$ the integral $I^{\delta }(\xi
)=\pi \exp (-\xi ^{2}),$ which is a Gaussian function of $\xi $. We take the
broadening parameter $\delta $ = $10^{-2}\omega _{0}$ throughout this paper.
This value corresponds to realistic electron mobilities due to ripplon
scattering. \cite{SHS95,kov}

In Fig. 1 we show $S_{sp}(q,\omega )$ as a function of $\omega $ for
different values of the wavevector $q$. Here, the temperature $T=0.4$ K, the
electron density $N_{e}=10^{4}$ cm$^{-1}$ and the holding field $E_{\perp
}=3\times 10^{3}$ V/cm. The energy gap between the quantized levels at this
holding field is $\omega _{0}\simeq 0.784$ K which leads to electron
occupation up to the second and the third subbands of about 14\% and\ 2\%,
respectively, in relation to the first one. For the sake of consistency, we
plot $S_{sp}(q,\omega )$ in two parts: $S_{sp}^{{\rm even}}(q,\omega )=$ $%
-\sum_{j+j%
%TCIMACRO{\UNICODE[m]{0xb4}}%
%BeginExpansion
{\acute{}}%
%EndExpansion
={\rm even}}{\rm Im}\left[ \Pi _{jj^{\prime }}^{\delta }(q,\omega )\right] $
(solid curves) and $S_{sp}^{{\rm odd}}(q,\omega )=$ $-\sum_{j+j%
%TCIMACRO{\UNICODE[m]{0xb4}}%
%BeginExpansion
{\acute{}}%
%EndExpansion
={\rm odd}}{\rm Im}\left[ \Pi _{jj^{\prime }}^{\delta }(q,\omega )\right] $
(dashed curves).\ For small wavevector $q,$ the spectral weight of the
intersubband SPEs are mostly described by Gaussian functions centered at
energies $\omega =n\omega _{0}$, where $n=1,2,3,....$. On the other hand,
the intrasubband SPE are represented by the less weighted peak seen in the
lower energy part of the spectra. This peak is mostly due to excitations in
the first subband. Fig. 1 show us that SPEs%
%TCIMACRO{\UNICODE{0xb4} }%
%BeginExpansion
\'{}%
%EndExpansion
induced effects should be significant at $\omega =n\omega _{0}$ for small $q$%
.

We show in Fig. 2 the spectral weight (a) $S^{{\rm even}}(q,\omega )$ and
(b) $S^{{\rm odd}}(q,\omega )$ for different electron densities $N_{e}$ at a
small fixed wavevector $q=0.1k_{0}$. The temperature is taken as $T$ $=0.4$
K. The density varies from $N_{e}$ $=1\times 10^{3}$ cm$^{-1}$ (the lowest
curve) until $N_{e}$ $=1.45\times 10^{4}$ cm$^{-1}$ (the top curve) with a
difference $1.5\times 10^{3}$ cm$^{-1}$. For $N_{e}=1\times 10^{3}$cm$^{-1}$%
, only the first subband has a significant electron density so the plasmon
modes related to this subband can be observed. This help us to identify the
different peaks corresponding to the intrasubband mode $(1,1)$ and
intersubband modes $(1,3)$, $(1,5)$, and $(1,7)$ in Fig. 2(a). The peaks in
Fig. 2(b), at the same electron density, are due to the plasmon modes $(1,2)$%
, $(1,4)$, and $(1,6)$. As the density increases, these peaks shift to
higher energy because the depolarization shift is enhanced. Meanwhile, the
intersubband modes related to the second subband appear as indicated in the
figure. We also observe that the Landau damping becomes stronger when the
plasmon mode $(1,n)$ approaches to the frequency $(n+1)\omega _{0}$ where
the single-particle excitations are of high intensity as shown in Fig. 1.
The insets show the energy position of the peaks due to the intersubband
plasmons as a function of the density $N_{e}$. The gaps result from the
Landau damping and are clearly seen around the energies $\omega =4\omega
_{0} $ and $6\omega _{0}$ in Fig. 2(a); and $\omega =3\omega _{0}$ and $%
5\omega _{0}$ in Fig. 2(b).

In Fig. 3, we plot the spectral weight function for different $q$ values
corresponding to those in Fig. 1 with $N_{e}=10^{4}$cm$^{-1}$ and $T=0.4$ K.
Figs. 3(a) and 3(b) show $S^{{\rm even}}(q,\omega )$ and $S^{{\rm odd}%
}(q,\omega ),$ respectively. The dispersion relations of the plasmon modes
obtained from the peak position are given in the insets. We also see small
peaks in the full spectra due to single-particle excitations indicated by
the open-dots in the insets. The lowest branch (open-dots) represents the
intrasubband SPEs. Similarly as in Q1D Fermi liquid electron systems at zero
temperature, with increasing $q$, the spectral weight of the intrasubband
(intersubband) plasmon mode increases (decreases). We also observe the
Landau damping induced gaps appearing around $\omega =n\omega _{0}$ at small 
$q$ where the SPEs are strong. Furthermore, the peaks due to the plasmon
excitations become wider and lower at larger $q$ because $S_{sp}(q,\omega )$
tends to be a more uniform function of $\omega $.

For the sake of completeness, we analyze in Fig. 4 the spectral weight $%
S(q,\omega )$ as a function of the energy $\omega $ and for several
temperatures. Here, the density $N_{e}$ $=10^{4}$ cm$^{-1}$ and $q=0.1k_{0}$%
. From the bottom to the top curves, the temperature increases from $T=0.2$
K to $2.0$ K with a step of $0.2$ K. As the temperature increases, the peaks
of the intersubband plasmon modes $(1,n)$ related to the first subband shift
to lower frequency while those modes $(2,n)$ related to the second subband
shift to higher frequency. Such a change is mainly induced by the
redistribution of the electron density in different subbands. As temperature
increases, the electron density of the first subband decreases but that of
the higher subband increases. The decrease of the electron density in the
first subband results in a decrease of the depolarization shift of the
related intersubband plasmon modes $(1,n)$. The insets show the energy
position of these peaks as a function of temperature.

In summary, we investigated the plasmon modes of the Q1D electron systems on
the surface of the liquid helium. We used a multisubband approach and
treated the system as a classical nondegenerate gas obeying
Maxwell-Boltzmann statistics. We found strong intersubband plasmon modes
related to the first two subbands. The single-particle excitations in the
system is responsible for strong Landau damping at frequencies $\omega
=n\omega _{0}$ where gaps appear in the plasmon spectra.

This work is supported by {\em FAPESP} (the research foundation agency of
the state of S\~{a}o Paulo), CNPq (Brazil), IUAP-V, GOA (Antwerp, Belgium)
and the European Community's Human Potential Programme under contract
HPRN-CT-2000-00157 \ ``Surface electrons''.

%\vspace{7.0cm}

\begin{figure}[tbp]
\caption{Spectral weight of the single-particle excitations $S_{sp}^{{\rm %
even}}(q,\protect\omega )$ (solid curves) and $S_{sp}^{{\rm odd}}(q,\protect%
\omega )$ (dashed curves) for $T =0.4$ K and $N_e=10^4$ cm$^{-1}$ with
different wavevectors q as indicated. }
\label{fig1}
\end{figure}

\begin{figure}[tbp]
\caption{The spectral weight (a) $S^{{\rm even}}(q,\protect\omega )$ and (b) 
$S^{{\rm odd}}(q,\protect\omega )$ for $q=0.1k_{0}$ and $T$ $=0.4$ K. The
electron densities are from $N_{e}$ $=1\times 10^{3}$ cm$^{-1}$ (the lowest
curve) to $N_{e}=1.45\times 10^{4}$ cm$^{-1}$ (the top curve) with a
difference $1.5\times 10^{3}$ cm$^{-1}$. The insets indicate the position of
the peaks as a function of $N_{e}$. }
\label{fig2}
\end{figure}

\begin{figure}[tbp]
\caption{The spectral weight (a) $S^{{\rm even}}(q,\protect\omega )$ and (b) 
$S^{{\rm odd}}(q,\protect\omega )$ for $T=0.4$ K, $N_{e}=10^{4}$ cm$^{-1}$,
and different wavevectors corresponding to those in Fig. 1. The insets give
the dispersion relation of the plasmon modes. The open-dots indicate the
small peaks due to single-particle excitations}
\label{fig3}
\end{figure}

\begin{figure}[tbp]
\caption{The spectral weight (a) $S^{{\rm even}}(q,\protect\omega )$ and (b) 
$S^{{\rm odd}}(q,\protect\omega )$ for q = 0.1 $k_{0}$ and $N_{e}=10^{4}$ cm$%
^{-1}$. The temperature increases from $T=0.2$ K (the lowest curve) to $2.0$
K (the top curve) with a step of $0.2$ K. The insets show the position of
the peaks. }
\label{fig4}
\end{figure}

\widetext

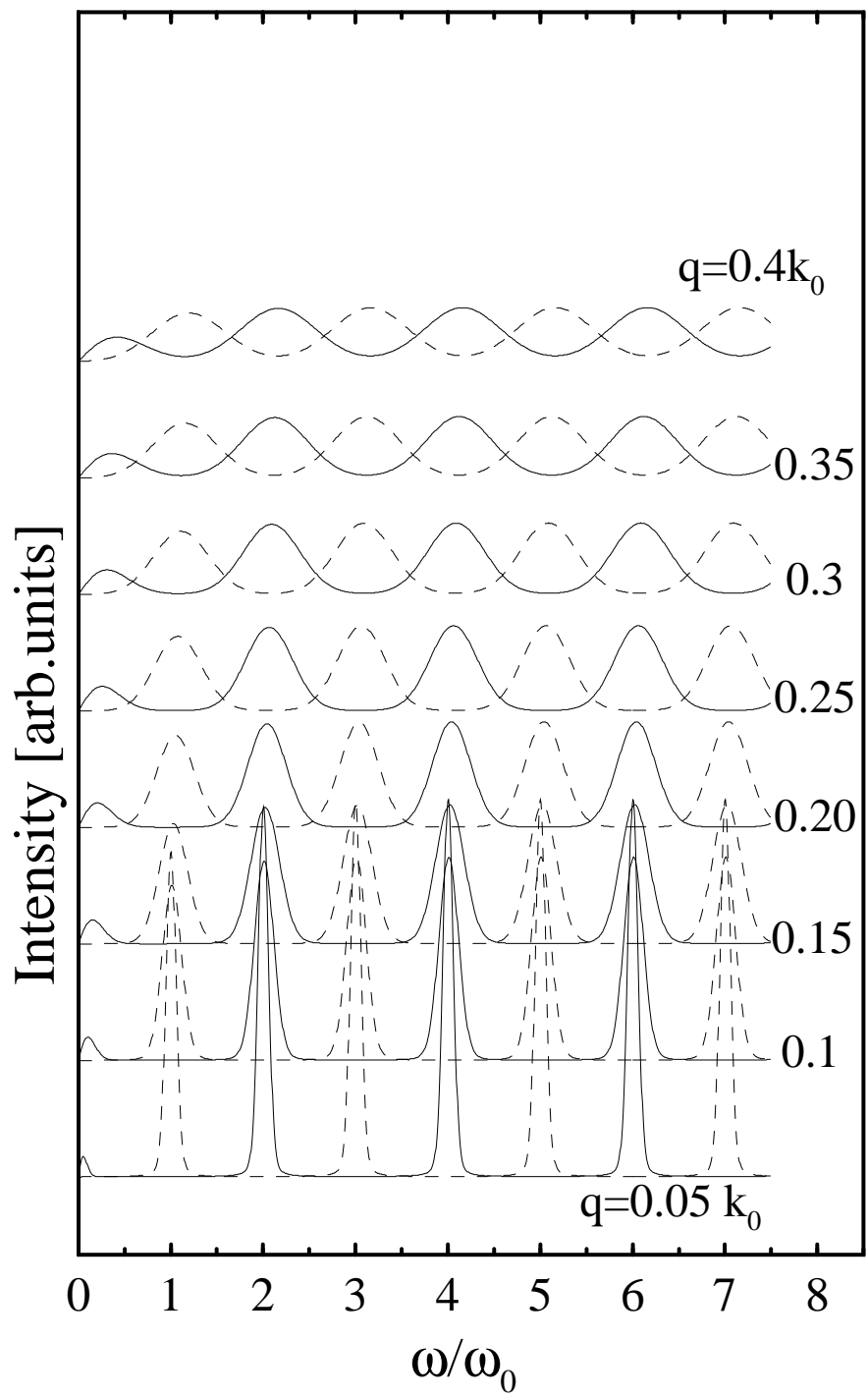
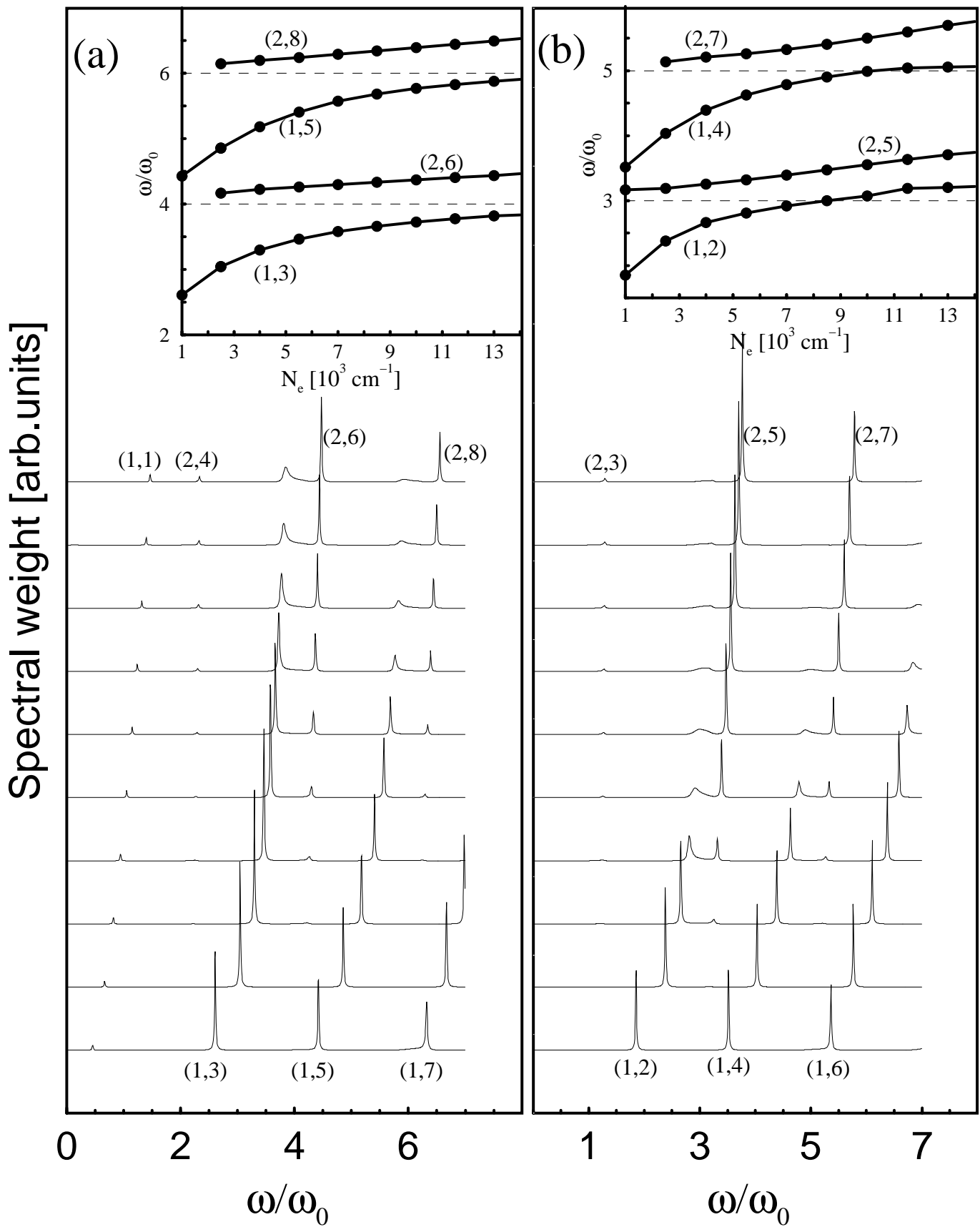
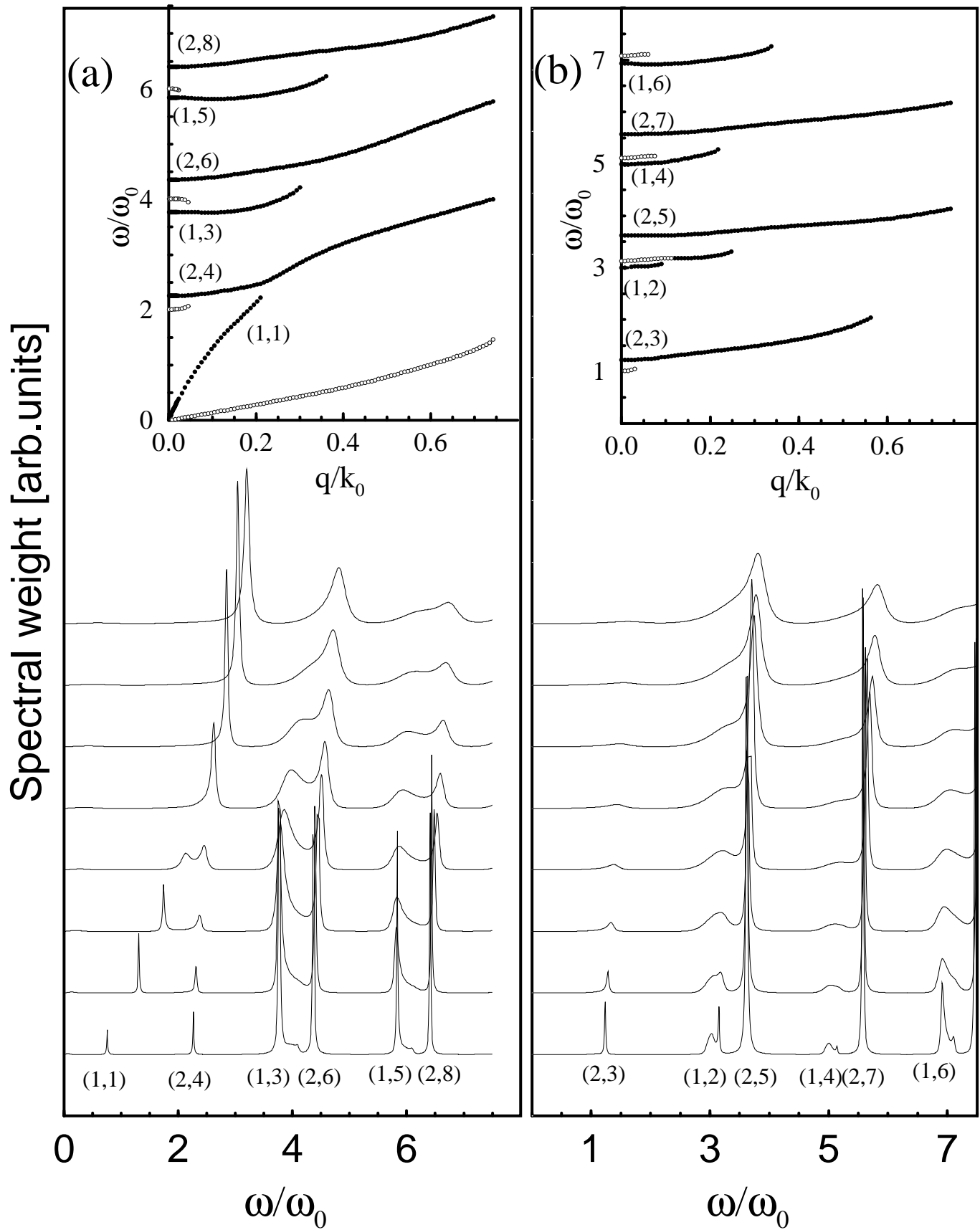
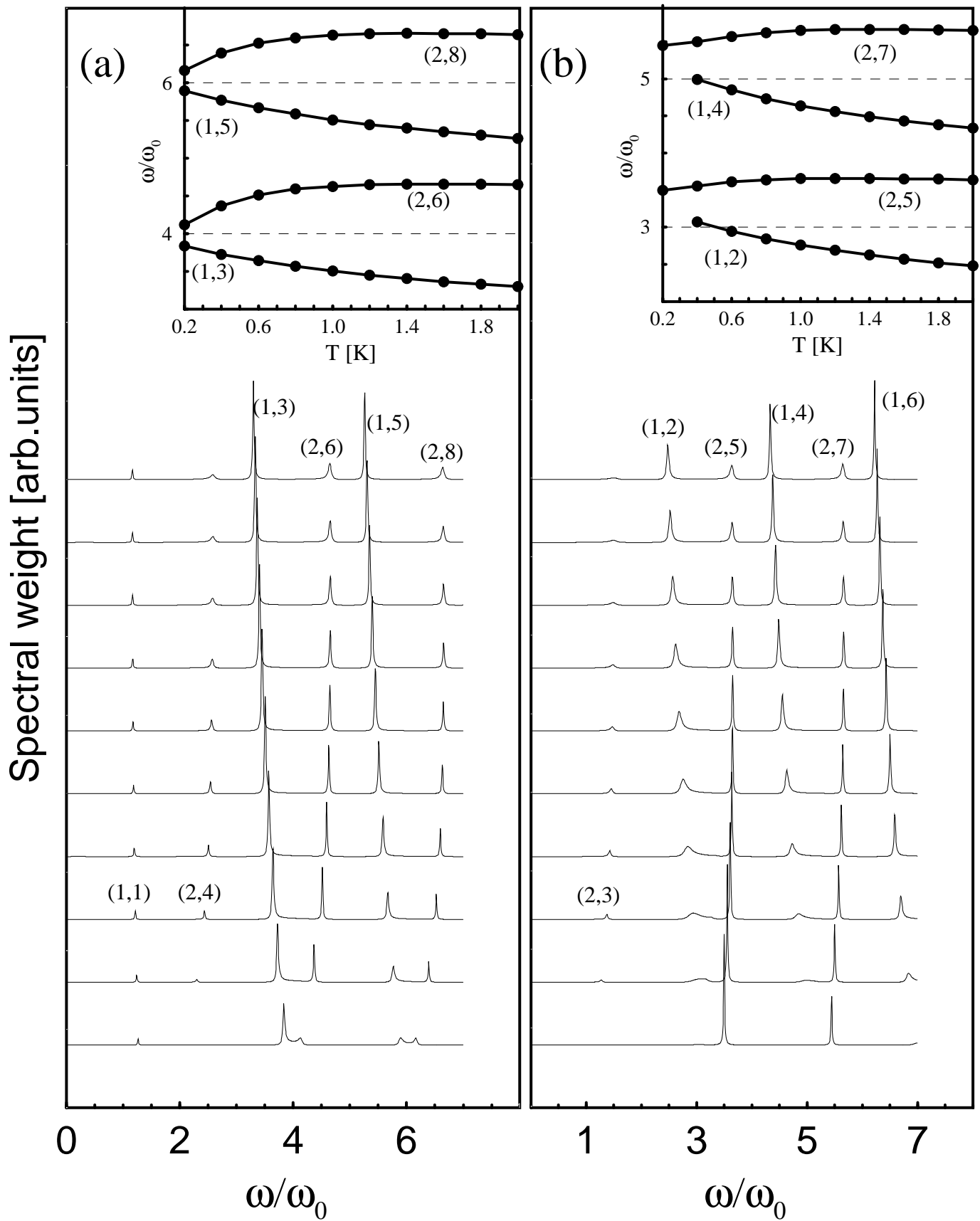
\end{document}